\documentclass[a4paper,english,10pt,twoside]{article}
\usepackage[english]{babel}
\usepackage[latin1]{inputenc}
\usepackage{amsmath}
\usepackage{graphicx}

\usepackage{fancyhdr}
\fancypagestyle{plain}{%
  \fancyhf{}%
  \fancyhead[C]{\small\it{Twenty years of giant exoplanets - Proceedings of the Haute Provence Observatory Colloquium, 5-9 October 2015\\ Edited by I. Boisse, O. Demangeon, F. Bouchy \& L. Arnold} }
  \fancyfoot[C]{\normalsize \thepage}%
}

\newcommand{\nc}   {\newcommand}
\nc{\dsfrac}[2]{{\displaystyle\frac{#1}{#2}}}

\newcommand{\cm}    {~\mathrm{cm}}

\newcommand{\Jcal}  {{\EuScript J}}

\nc{\ergs}  {\mathrm{erg}/\mathrm{gm\;s}}
\nc{\ergc}  {{\rm erg}~{\rm cm}^{-3}}
\nc{\ergcs} {{\rm erg}~{\rm cm}^{-2}~{\rm s}^{-1}}
\nc{\ms}   {{\rm m}~{\rm s}^{-1}}

\nc{\Eem}   {\widetilde{E}_\mathrm{e}}
\nc{\Eexm}  {\widetilde{E}_\mathrm{ex}}
\nc{\Eim}   {\widetilde{E}_\mathrm{I}}
\nc{\ER}    {E_\mathrm{R}}
\nc{\FL}    {F_\mathrm{Ly}}
\nc{\FR}    {F_\mathrm{R}}
\nc{\FRa}   {F_\mathrm{R,1}}
\nc{\FFR}   {\mathbf{F}_\mathrm{R}}
\nc{\Fe}    {F_\mathrm{e}}
\nc{\FFe}   {\mathbf{F}_\mathrm{e}}
\nc{\Fsat}  {q_\mathrm{sat}}
\nc{\Ke}    {\EuScript{K}_\mathrm{e}}
\nc{\mfp}   {l_\mathrm{e}}
\nc{\mrc}   {\mathrm{c}}
\nc{\mrr}   {\mathrm{r}}
\nc{\nH}    {n_\mathrm{H}}
\nc{\nel}   {n_\mathrm{e}}
\nc{\Pe}    {P_\mathrm{e}}
\nc{\Pg}    {P_\mathrm{g}}
\nc{\PR}    {P_\mathrm{R}}
\nc{\Qelc}  {Q_\mathrm{elc}}
\nc{\Qinc}  {Q_\mathrm{inc}}
\nc{\gyr}   {r_\mathrm{B}}
\nc{\Ta}    {T_a}

\nc{\divFe} {\nabla\cdot\FFe}
\nc{\divFR} {\nabla\cdot\FFR}
\nc{\FLj}   {F_\mathrm{Ly,\Jcal}}
\nc{\FRj}   {F_\mathrm{R,\Jcal}}

\nc{\aap}   {A\&A}
\nc{\aapr}  {A\&ARv}
\nc{\aaps}  {A\&AS}
\nc{\apj}   {ApJ}
\nc{\icarus}{Icarus}
\nc{\mnras} {MNRAS}
\nc{\nat}   {Nature}
\nc{\pasp}  {PASP}

\usepackage{txfonts}
\usepackage[squaren,cdot]{SIunits}
\usepackage{tabularx}
\usepackage[breaklinks=true]{hyperref} 
\usepackage{natbib,twoopt}
\bibpunct{(}{)}{;}{a}{}{,} 
\makeatletter
\newcommandtwoopt{\citeads}[3][][]{\href{http://adsabs.harvard.edu/abs/#3}%
{\def\hyper@linkstart##1##2{}%
\let\hyper@linkend\@empty\citealp[#1][#2]{#3}}}
\newcommandtwoopt{\citepads}[3][][]{\href{http://adsabs.harvard.edu/abs/#3}%
{\def\hyper@linkstart##1##2{}%
\let\hyper@linkend\@empty\citep[#1][#2]{#3}}}
\newcommandtwoopt{\citetads}[3][][]{\href{http://adsabs.harvard.edu/abs/#3}%
{\def\hyper@linkstart##1##2{}%
\let\hyper@linkend\@empty\citet[#1][#2]{#3}}}
\newcommandtwoopt{\citeyearads}[3][][]%
{\href{http://adsabs.harvard.edu/abs/#3}
{\def\hyper@linkstart##1##2{}%
\let\hyper@linkend\@empty\citeyear[#1][#2]{#3}}}
\makeatother

\makeatletter 

\def\@maketitle{%
  \vskip 2em%
  \begin{center}%
  \let \footnote \thanks
    {\LARGE\textbf \@title \par}%
    \vskip 1.5em%
    {\normalsize
      \lineskip .5em%
      \begin{tabular}[t]{c}%
        \@author
      \end{tabular}\par}%
    \vskip 1em%
    {\normalsize \@date}%
  \end{center}%
  \par
  \vskip 1.5em}
\makeatother

\newcommand{\affil}[1]{\small{\hskip-0.55cm #1}}

\begin{document}

\setcounter{page}{1}  

\title{How to form asteroids from mm-sized grains} 
\author{D.\,Carrera$^{1}$, A.\,Johansen$^{1}$, M.\,B.\,Davies$^{1}$} 
\date{} 
\maketitle
\affil{ $^1$Lund Observatory,
        Dept. of Astronomy and Theoretical Physics,
        Lund University, Box 43, 22100 Lund, Sweden }


\begin{abstract}
The size distribution of asteroids in the solar system suggests that they formed top-down, with $100-1000$ km bodies forming from the gravitational collapse of dense clumps of small solid particles. We investigate the conditions under which solid particles can form dense clumps in a protoplanetary disc. We used a hydrodynamic code to model the solid-gas interaction in disc. We found that particles down to millimeter size can form dense clumps, but only in regions where solids make $\sim$8\% of the local surface density. More generally, we mapped the range of particle sizes and concentrations that is consistent with the formation of particle clumps.
\end{abstract}

\section{Introduction}

Planetesimals are 10-1000 km bodies that form the seeds of terrestrial planets, as well as the cores of ice giants and gas giants \citep[e.g.][]{Safronov_1972, Chiang_2010, Johansen_2014}. Large asteroids are left-over planetesimals that were never incorporated into planets. The largest asteroid in the solar system is Vesta, with a diameter of $\sim$500 km \citep{Russell_2012}. The size distribution of asteroids suggests that planetesimals form in a top-down process, where bodies larger than 100 km formed first, and smaller ones formed later by collisional grinding \citep{Morbidelli_2009}. In this scenario, asteroids would form from the gravitational collapse of a large clump of smaller particles. One way to produce this kind of concentration is the streaming instability \citep{Youdin_2005}. It is driven by the relative drift between the solid and gas components of the disk. The streaming instability has already proven effective in forming planetesimals in simulations with 10-100 cm sized particles \citep{Johansen_2007, Johansen_2007b, Bai_2010b}. However, asteroids are made of particles much smaller than 10 cm. Most of the mass in primitive meteorites consists of small $\sim$1 mm particles called \textit{chondrules} \citep[e.g.][]{Jacquet_2014}. \citet{Ormel_2008} have shown that chondrules may be able to form aggregates a few millimeters in size, and that weak turbulence allows for larger aggregates.

In \citet{Carrera_2015} we establish the connection between the streaming instability and chondrules. We probe the limits of the streaming instability at both the small-particle and large-particle limits. We show that particles a few millimeters in size (e.g. chondrule aggregates) can form particle clumps inside a protoplanetary disc.

\section{Methods}

We use the Pencil Code \citep{Youdin_2007} to model a two-dimensional (azimuthally symmetric) slice of a protoplanetary disc. We follow the canonical model for the disc where the solar system formed, known as the minimum mass solar nebula \citep[MMSN,][]{Hayashi_1981}. In it, the surface density of the gas component of the disk follows the power law

\begin{equation}\label{eqn:hayashi}
    \Sigma = 1700 \, {\rm g} \, {\rm cm}^{-2} \, \left( \frac{r}{\rm AU} \right)^{-3/2}.
\end{equation}

In addition to the gas, the solid component of the disc represents 1\% of the initial disc mass, and follows the same power law. The gas is modelled in a $128 \times 128$ square grid, while the solids are modelled as particles. The total size of our box is $0.2 \times 0.2$ times the disc scale height. The behaviour of particles in the disc is driven largely by the particle friction time $t_{\rm f}$. That is the time needed for gas drag to change the velocity of a particle by order unity. For particles smaller than the mean free path of gas particles, the friction time is given by 

\begin{equation}
    t_{\rm f} \, = \, \frac{\rho_\bullet \, R}{\rho \, c_{\rm s}} \sqrt{\frac{\pi}{8}},
\end{equation}
where $\rho$ is the gas density, $\rho_\bullet$ is the material density of the solid particles, $R$ is the radius of the particle, and $c_{\rm s}$ is the sound speed. The \textit{Stokes number} of a particle ($\tau_{\rm f}$) is the friction time expressed in terms of the Keplerian frequency. The other key parameters are the speed of the headwind experienced by solid particles, in terms of the sound speed ($\Delta$), and the solid concentration ($Z$).

\begin{eqnarray*}
    \tau_{\rm f} \, &=& \, t_{\rm f} \, \Omega_{\rm k}
            \, = \, \frac{\rho_\bullet \, R}{\rho \, c_{\rm s}}
            \, \Omega_{\rm k} \sqrt{\frac{\pi}{8}}, \\
    \Delta \, &=& \, \frac{v_{\rm solid} - v_{\rm gas}}{c_{\rm s}}, \\
    Z \, &=& \, \frac{\Sigma_{\rm solid}}{\Sigma_{\rm gas} + \Sigma_{\rm solid}}.
\end{eqnarray*}
For the minimum mass solar nebula, $\Delta = 0.05$ and $Z \sim 0.01$. In our simulations we use $\Delta = 0.05$ but we start with $Z = 0.005$. For each run we fix the value of $\tau_{\rm f}$, and we gradually increase $Z$ and observe at which moment the solid particles form visible clumps. We tested 17 particle sizes from $\tau_{\rm f} = 10^{-3}$ to $\tau_{\rm f} = 10$, and for each particle size we performed three runs with different (random) initial distributions of particles.

\section{Results and discussion}

\begin{figure}[hb!]
    \centering
    \includegraphics[width=0.7\textwidth]{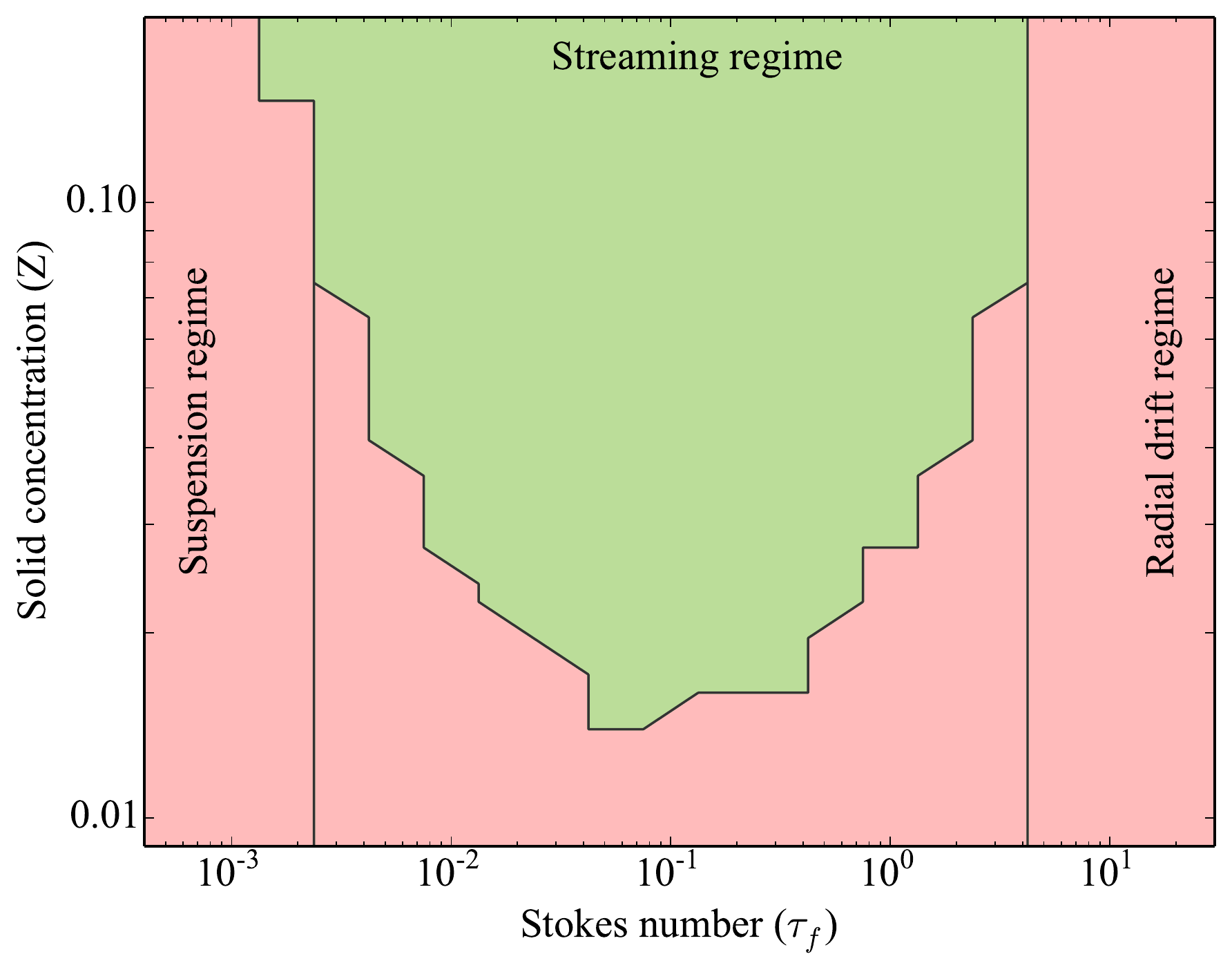}
    
    \caption{The green region marks the range of particle concentrations ($Z$) and particle size ($\tau_{\rm f}$) where particle clumps form by the streaming instability. We also find that for $\tau_{\rm f} \lesssim 0.002$ particles are suspended in the midplane with no appreciable radial drift. For $\tau_{\rm f} \gtrsim 4$, radial drift is too rapid to allow particles to clump. This figure is a simplified version of Fig.\,8 from \citep{Carrera_2015}.}
    \label{fig:key-results}
\end{figure}

Our key results are shown in Fig.\,\ref{fig:key-results}. The green region marks the range of particle sizes ($\tau_{\rm f}$) and concentrations ($Z$) where our simulations show visible particle clumps. The least squares fit for this region is

\begin{equation}\label{eqn:key-equation}
    \log_{\rm 10}(Z_{\rm crit}) \, = \, -1.86
                + 0.3 \left( \log_{\rm 10}(\tau_{\rm f}) + 0.98 \right)^2,
\end{equation}
where particle clumps occur for $Z > Z_{\rm crit}$. The conversion factor $R \sim 78 \cm \; \tau_{\rm f}$ is valid for the MMSN at 2.5 AU.

\begin{figure}[hb!]
    \centering
    \includegraphics[width=0.65\textwidth]{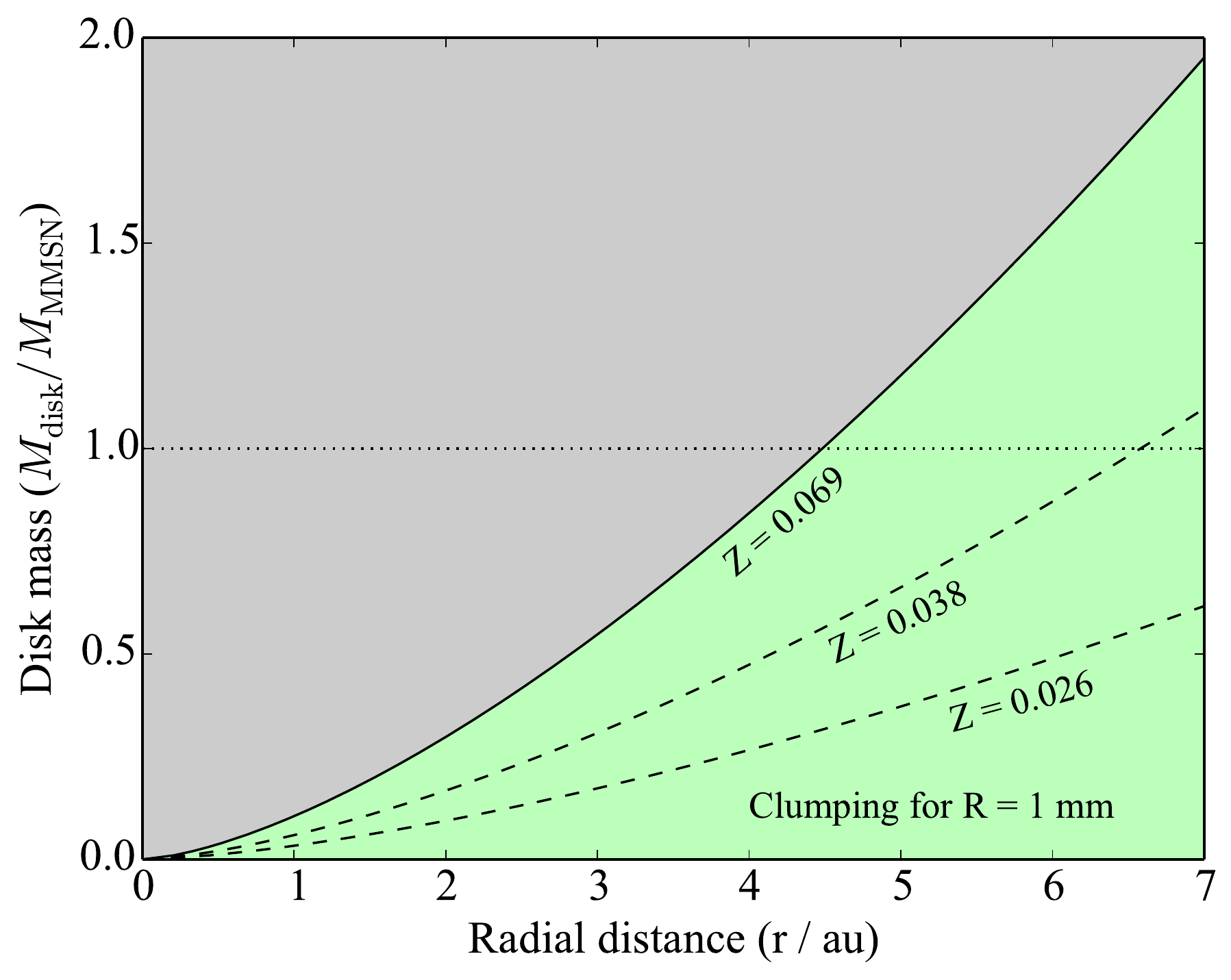} \\
    \includegraphics[width=0.65\textwidth]{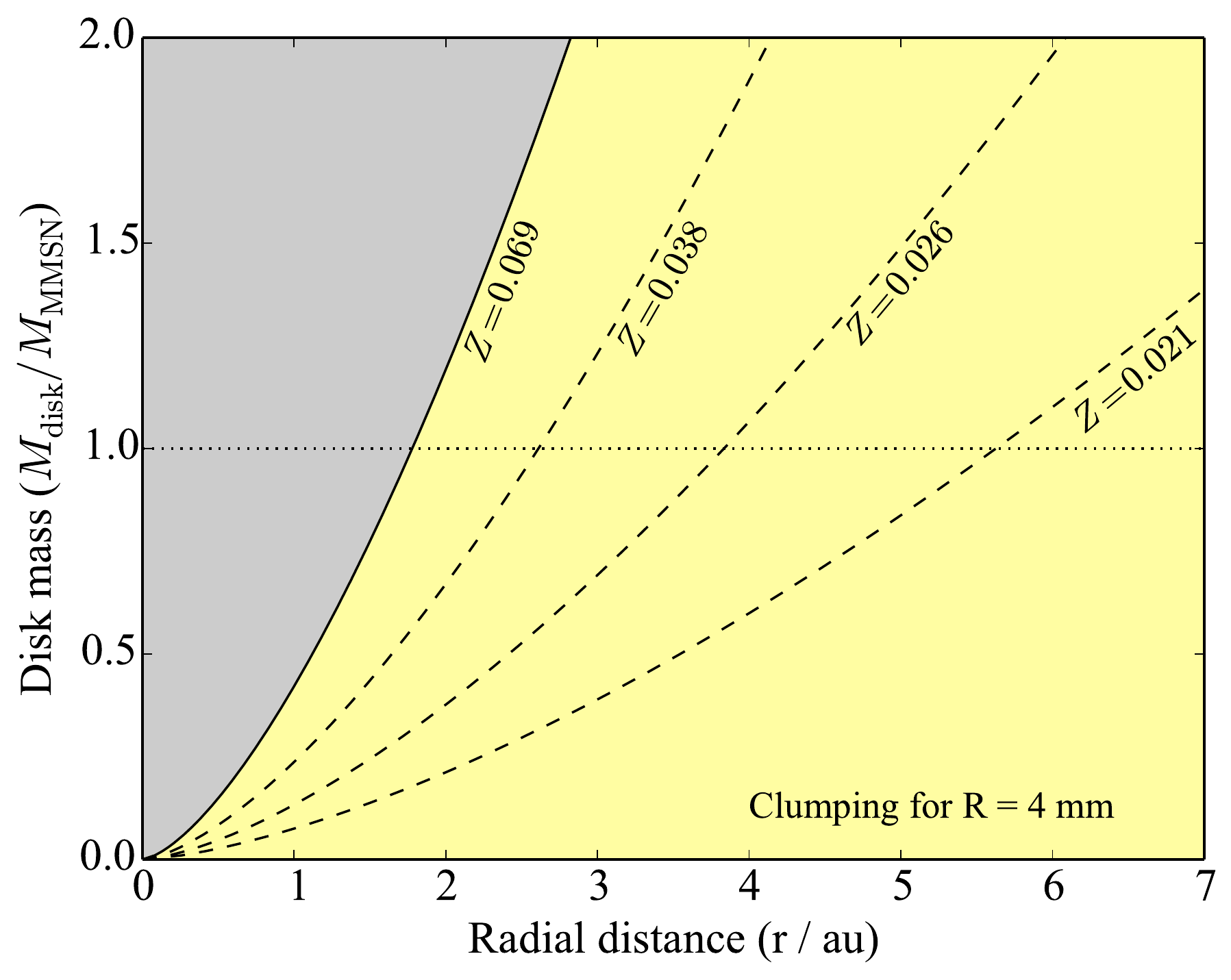}
    
    \caption{Region of the protoplanetary disc where particle clumps can form. The disc mass is expressed in terms of the minimum mass solar nebula, and corresponds to the disc evolution. Larger values of $R$ and $Z$ bring the asteroid-forming region closer to the star. In all cases, it is difficult to form planetesimals closer than 1.5 AU until late in the disc evolution. Figure reproduced from \citep{Carrera_2015}.}
    \label{fig:application}
\end{figure}

In the context of forming asteroids at 2.5 AU, the key take-away message from Fig.\,\ref{fig:key-results} is that it is possible for particles as small as $R \sim 2$ mm to form particle clumps through the streaming instability. However, this comes at the expense of a very high particle concentration ($Z \sim 0.08$) that may be unfeasible for a protoplanetary disc. Our results also show that a small increase in particle size, above the $R \sim 2$ mm minimum, significantly reduces the particle concentration required to make clumps. Therefore, a realistic model of asteroid formation may require either some amount of gas dispersal (e.g. disc winds or photoevaporation) or additional coagulation before asteroids can form. This point is also illustrated in Fig.\,\ref{fig:application}, where we show the region of the disc where asteroids can form as a function of disc mass, for different choices of $R$ and $Z$.

\section{Conclusions}

The formation of asteroids remains a difficult problem. We have completed over 100 computer simulations where we modelled the dynamics of solid particles inside a protoplanetary disc. We probe a wide range of particle sizes and concentrations to test the limits of the streaming instability. Our key results are:

\begin{itemize}

\item We present the range of particle sizes and concentrations that are consistent with the formation of clumps by the streaming instability. The region is summarized in Fig.\,\ref{fig:key-results} and Eq.\,\ref{eqn:key-equation}.

\item We find that particles as small as $\tau_{\rm f} \sim 10^{-3}$ ($R \sim$2 mm at 2.5 AU) can form particle clumps. However, this only occurs at very large particle concentrations ($Z \sim 0.08$) that may not occur in nature. The required concentration drops rapidly for $\tau_{\rm f}$ slightly higher than the minimum.

\end{itemize}

Altogether we find that particle concentration by the streaming instability provides a viable path to forming asteroids directly from mm-sized chondrules, particularly if weak turbulence allows for larger chondrule aggregates to form.

\section{Acknowledgements}

We acknowledge the support from the Knut and Alice Wallenberg Foundation, the Swedish Research Council (grants 2010-3710 and 2011-3991) and the European Research Council Starting Grant 278675-PEBBLE2PLANET that made this work possible. Computer simulations were performed using the Alarik cluster at Lunarc Center for Scientific and Technical Computing at Lund University. Some simulation hardware was purchased with grants from the Royal Physiographic Society of Lund.


\bibliography{main}

\begin{thebibliography}{14}
\expandafter\ifx\csname natexlab\endcsname\relax\def\natexlab#1{#1}\fi

\bibitem[{{Bai} \& {Stone}(2010)}]{Bai_2010b}
{Bai}, X.-N. \& {Stone}, J.~M. 2010, \apj, 722, 1437

\bibitem[{{Carrera} {et~al.}(2015){Carrera}, {Johansen}, \&
  {Davies}}]{Carrera_2015}
{Carrera}, D., {Johansen}, A., \& {Davies}, M.~B. 2015, \aap, 579, A43

\bibitem[{{Chiang} \& {Youdin}(2010)}]{Chiang_2010}
{Chiang}, E. \& {Youdin}, A.~N. 2010, Annual Review of Earth and Planetary
  Sciences, 38, 493

\bibitem[{{Hayashi}(1981)}]{Hayashi_1981}
{Hayashi}, C. 1981, Progress of Theoretical Physics Supplement, 70, 35

\bibitem[{{Jacquet}(2014)}]{Jacquet_2014}
{Jacquet}, E. 2014, \icarus, 232, 176

\bibitem[{{Johansen} {et~al.}(2014){Johansen}, {Blum}, {Tanaka}, {Ormel},
  {Bizzarro}, \& {Rickman}}]{Johansen_2014}
{Johansen}, A., {Blum}, J., {Tanaka}, H., {et~al.} 2014, ArXiv e-prints

\bibitem[{{Johansen} {et~al.}(2007){Johansen}, {Oishi}, {Mac Low}, {Klahr},
  {Henning}, \& {Youdin}}]{Johansen_2007}
{Johansen}, A., {Oishi}, J.~S., {Mac Low}, M.-M., {et~al.} 2007, \nat, 448,
  1022

\bibitem[{{Johansen} \& {Youdin}(2007)}]{Johansen_2007b}
{Johansen}, A. \& {Youdin}, A. 2007, \apj, 662, 627

\bibitem[{{Morbidelli} {et~al.}(2009){Morbidelli}, {Bottke}, {Nesvorn{\'y}}, \&
  {Levison}}]{Morbidelli_2009}
{Morbidelli}, A., {Bottke}, W.~F., {Nesvorn{\'y}}, D., \& {Levison}, H.~F.
  2009, \icarus, 204, 558

\bibitem[{{Ormel} {et~al.}(2008){Ormel}, {Cuzzi}, \& {Tielens}}]{Ormel_2008}
{Ormel}, C.~W., {Cuzzi}, J.~N., \& {Tielens}, A.~G.~G.~M. 2008, \apj, 679, 1588

\bibitem[{{Russell} {et~al.}(2012){Russell}, {Raymond}, {Coradini}, {McSween},
  {Zuber}, {Nathues}, {De Sanctis}, {Jaumann}, {Konopliv}, {Preusker}, {Asmar},
  {Park}, {Gaskell}, {Keller}, {Mottola}, {Roatsch}, {Scully}, {Smith},
  {Tricarico}, {Toplis}, {Christensen}, {Feldman}, {Lawrence}, {McCoy},
  {Prettyman}, {Reedy}, {Sykes}, \& {Titus}}]{Russell_2012}
{Russell}, C.~T., {Raymond}, C.~A., {Coradini}, A., {et~al.} 2012, Science,
  336, 684

\bibitem[{{Safronov}(1972)}]{Safronov_1972}
{Safronov}, V.~S. 1972, {Evolution of the protoplanetary cloud and formation of
  the earth and planets.} (Keter Publishing House)

\bibitem[{{Youdin} \& {Johansen}(2007)}]{Youdin_2007}
{Youdin}, A. \& {Johansen}, A. 2007, \apj, 662, 613

\bibitem[{{Youdin} \& {Goodman}(2005)}]{Youdin_2005}
{Youdin}, A.~N. \& {Goodman}, J. 2005, \apj, 620, 459

\end{thebibliography}
\bibliographystyle{aa} 

\end{document}